# Micromagnetic Modeling of Telegraphic Mode Jumping in Microwave Spin Torque Oscillators

B. Gunnar Malm
*School of EECS*
KTH Royal Institute of Technology
Kista, Sweden
gunta@kth.se

Anders Eklund
*School of EECS*
KTH Royal Institute of Technology
Kista, Sweden
ajeklund@kth.se

Mykola Dvornik
*Department of Physics*
Gothenburg University
Gothenburg, Sweden
mykola.dvornik@physics.gu.se

*Abstract*—The time domain stability of microwave spin torque oscillators (STOs) has been investigated by systematic micromagnetic simulations. A model based on internal spin wave reflection at grain boundaries with reduced exchange coupling was implemented and used to study the oscillator under quasi-stable operating conditions. Telegraphic mode jumping between two operating frequencies (23.3 and 24.1 GHz) was observed in the time domain with characteristic dwell times in the range of 10-100 ns. The oscillating volume was shown to have a different shape at the distinct operating frequencies. The shape difference is governed by spin wave reflections at the grain boundaries. The resulting non-linear behavior of the oscillator was shown to be a collective effect of spin wave scattering at different locations within a few spin wavelengths from the nano-contact.

## I. Introduction

Spintronic devices are interesting alternatives for novel memory and logic functionality, and can also be used for microwave generation and detection. These device are based on a stack of thin magnetic layers and have a very small footprint, e.g. compared to advanced CMOS devices and hence they can be integrated into the back-end of-line processing. In spin torque oscillators (STOs) for microwave application the frequency is tuned by a DC current flowing into a nano-contact (NC) with diameter less than 100 nm. The current-frequency relation, or slope *df/dI*, typically shows non-linear regions of operation as well as separated modes or frequency jumps. The non-linear operation regions are accompanied by an increased linewidth (white noise) as well as pronounced low-frequency $1/f$ noise [1]. In our previous work [2] a model based on internal spin wave reflection, due to reduced exchange coupling at grain boundaries [3, 4], was proposed and could successfully reproduce a wide body of experimental observations. In this work, for the first time, we use systematic simulations in the GPU accelerated micromagnetic software MuMax$^3$ [5, 6] to study the time domain stability of the different frequency modes. The simulations included a stochastic thermal field (T = 300 K) that acts on the individual micromagnetic spins. A characteristic telegraphic switching was observed for DC currents that correspond to abrupt frequency jumps

## II. Micromagnetic simulation approach

Micromagnetic simulations are based on a finite-difference time domain numerical solution of the well-known LLGS equation including a spin transfer torque (STT) term. The equation is discretized onto a mesh where each node or cell of the mesh represents a single micro spin quantity. For typical simulation domain sizes GPU accelerated codes provide a significant simulation time reduction over CPU based codes. In our study, since we are considering propagation and reflection of spin waves in an extended magnetic film that is significantly larger than the actual oscillating volume located underneath the nano-contact (100 nm NC diameter vs. 640 nm square), see Fig. 1.

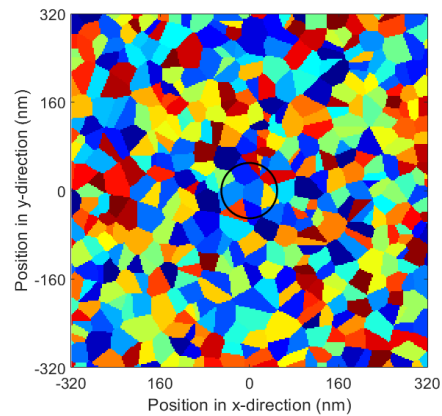

Fig. 1. Grain structures realized by a tesselation algorithm, average grain size 30 nm. The nominal placement and size of the nanocontact (NC) are indicated by the circle.

The simulation domain was divided into 256 by 256 cells in the x-y plane and 5 layers in the z-direction to represent the different layers of the stack. The layers correspond to our experimental devices with 8 nm Co (fixed layer), 8 nm Cu non-magnetic spacer, NiFe 4.5 nm (free layer). In this study we used a fixed NC size of 100 nm, an applied external field of 10 kOe, out-of-plane angle 70° and currents in the range 20.5 mA to 32.50 mA in order to excite the propagating mode. For these particular conditions the propagating mode becomes stable around 18 mA. As shown in Fig. 1 different grain structures can be implemented by providing a random seed to a tesselation algorithm. In addition to this the average exchange coupling at each individual grain boundary can also be randomized. For actual device processing the placement of the nano-contact with respect to the grain boundaries is arbitrary. In order to replicate this situation, in an eScience approach, we performed a set of simulations where the seed for grain generation was fixed while the nano-contact itself was displaced a few unit cells corresponding to either 5 or 12.5 nm in both x- and y-directions. The chosen displacement is smaller than the average grain size, 30 nm, and also smaller than the estimated spin wavelength (70-100 nm).

All simulations were performed for a duration of at least 100 ns and the solution, magnetization state $\mathbf{m} = (m_x, m_y, m_z)$ as well as the total energy density, $E_{tot} \propto \mathbf{m} \cdot \mathbf{B}_{eff}$, was stored for each increment of 5 ps. The internal time step, used by the adaptive solver, was observed to be in the range of $0.5 - 1 \times 10^{-13}$ s. Running simulations with a fixed internal time step of less than $1 \times 10^{-13}$ s yielded virtually identical results. Still the adaptive time stepping allows a reduction of total simulation time for cases with a finite temperature, random thermal field, turned on. The 100 ns time traces give a frequency resolution of 10 MHz. In order to get higher frequency resolution (1 MHz) and in particular to study long-term time domain stability we also performed many simulations of 1 μs duration. These finished in about 48 h on high performance GPUs (GTX-1080i, Tesla K80 or similar).

III. RESULTS AND DISCUSSION

As described in Sec. II we systematically varied the NC placement and the results of this set of simulations are summarized in Fig. 2. The power spectral density (PSD) of the y-component of the magnetization $m_y$ was obtained using the signal processing capability of MATLAB® and the frequency at peak power was extracted vs. DC current. Most of the placements result in a varying degree of continuous non-linear behavior while there is one case with a large mode jump (~1 GHz) and several with smaller jumps of a few 100 MHz. A reference case with full exchange coupling at all grain boundaries, replicating an ideal homogeneous material, indicates that the slope $df/dI$ is inherently constant for these oscillators. Such behavior is seldom observed in actual experimental devices, at least not for such a wide current sweep. In the following analysis we will focus on the time domain behavior of the mode transition at 27.75 mA for the (5, 5) case.

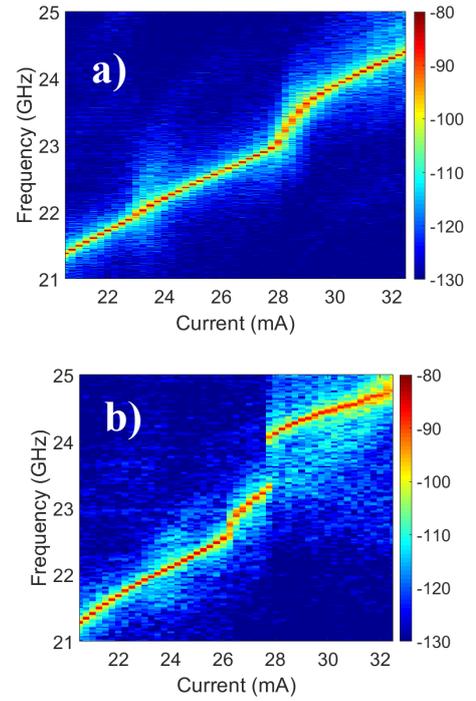

Fig. 3. Simulated output power spectral density (dB/Hz) vs. driving DC current. a) nominal placement of NC resulting in contiuous but non-linear behavior b) 12.5 nm offset in x- and y-direction showing two active frequency modes at 27.75 mA.

Fig. 4. shows the frequency spectra, centered around the mode transition, obtained with four different seeds to the pseudorandom thermal field generator. The two frequency peaks are reproducibly obtained, which rules out the possibility of a numerical instability as its origin, see also the time domain traces in Fig. 5. In all cases, two frequency modes at 23.3 and 24.1 GHz with comparable power are observed in the spectrum, meaning that the two modes are occurring with roughly the same probability over time, for this particular current. We have also observed cases where one of the peaks has considerably lower power, in that case the oscillator spends most of the time at one of the two possible frequencies.

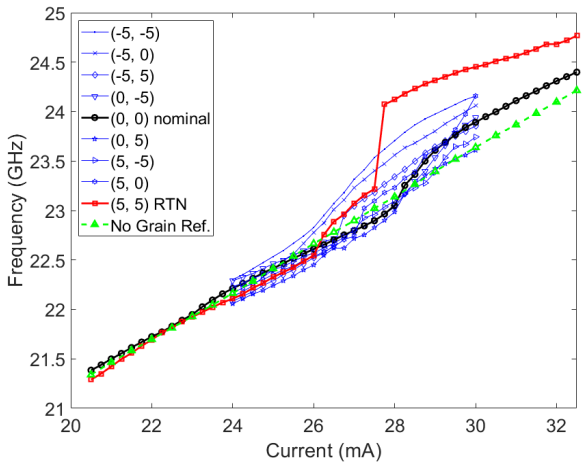

Fig. 2. Oscillation frequency at peak power for +/- 5 unit cell (+/- 12.5 nm) displacements of the nano-contact, a reference case with no grains is included..

In Fig.3 a) we illustrate the full PSD, both power and frequency, for the nominal case that shows a non-linearity and b) the RTN case that has both a non-linearity at lower current around 26 mA and clear jump at 27.75 mA. In the nominal case the spectrum clearly shows a broader linewidth in the regions with steeper slope $df/dI$ around 28 mA. The time domain data still shows a relatively slow frequency instability that does not exhibit any preferred low and high energy states that would yield a telegraphic behavior.

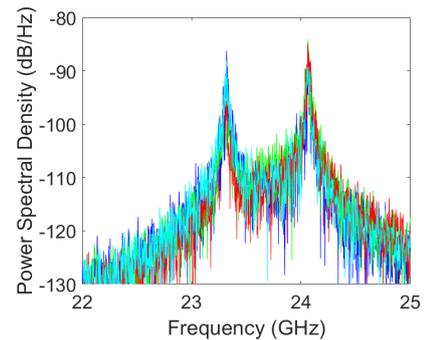

Fig. 4. Frequency spectra for different thermal field seeds showing two distinct peaks at 23.3 and 24.1 GHz at DC current 27.75 mA.

By plotting the total oscillator energy density vs. simulation time, shown in Fig. 5, it was found that the two frequency modes are excited in a telegraphic manner and hence do not co-exist in time. A moving average window of 0.5 ns was used to filter out faster energy fluctuations. Typical dwell times in the different modes are in the range of ten to hundred ns. For example the top panel of Fig. 5 shows two initial segments of 50 ns at the low and high energy states of the oscillator respectively. The longest observed stable segments are more than 200 ns in duration and are found at both energy levels. In our previous experimental study, we mainly focused on higher current where dwell times in the range tens of nanosecond and as high as milliseconds could be observed [7].

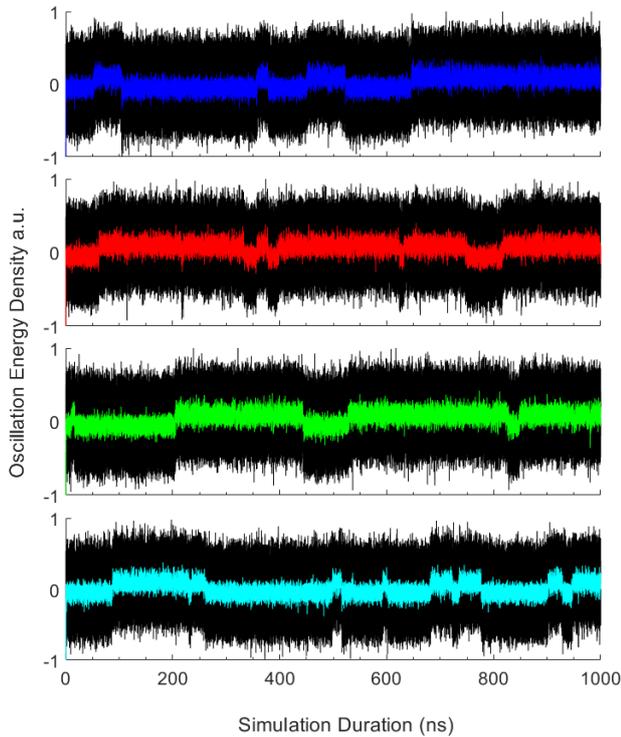

Fig. 5. Time domain representation of oscillator energy density at DC current 27.75 mA, for different thermal field seeds showing two mode telegraphic jumping, The colored lines are moving averages over 0.5 ns.

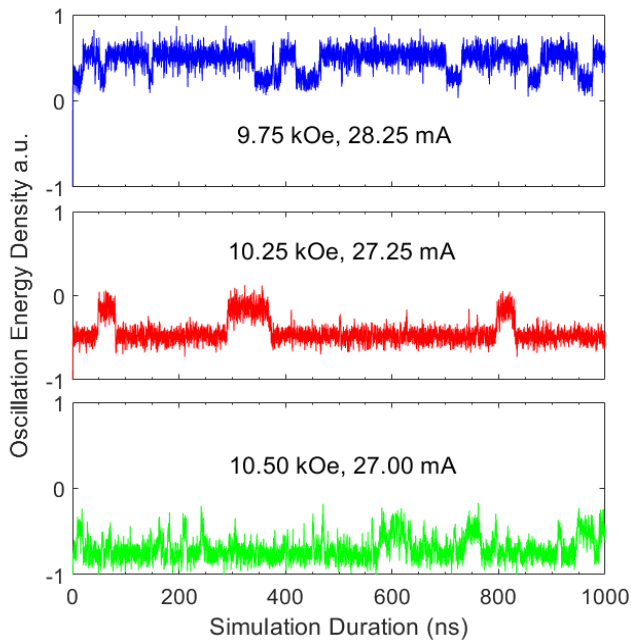

Fig. 6. Energy density (moving averages over 0.5 ns) for different applied fields and currents corresponding to the mode transitions between low and high oscillation frequencies..

The telegraphic behavior was relatively robust, changing the applied field in the range 9.75 – 11.0 kOeA, as well as using a lower temperature of 200 K still yielded two separated frequencies. In Fig. 6 the energy states are illustrated for three different applied fields, the current corresponding to the location of the frequency jump varies slightly. For the 9.75 kOe case the higher energy dominates, for the 10.25 kOe the lower energy dominates, while the 10.50 kOe case exhibits a clear lower energy and an indication of multiple possible higher energy states, with relatively short dwell times.

Any physical system that presents a telegraphic behavior must have some inherent energy barrier between the two states. To analyze this we performed a FFT over the micromagnetic data ($m_y$) in each mesh node. Similar results are obtained by analyzing the other magnetization components or the total energy. For example, the time trace with the first thermal seed showed that the oscillator operated at 23.3 GHz for the first 50 ns and at 24.1 GHz for the next 50 ns. The FFTs showed that the shape of the oscillating volume was distinctly different at these two frequencies, see Fig. 7 a) and b). The resulting shape of the oscillating volume is influenced by the interference of spin waves that are scattered or reflected at multiple grain boundaries, in particular at so called grain boundary triple junctions (corners). The oscillating volumes do not have a smooth spatial extent and there are regions (locations) indicated by the arrows where the oscillation is showing a standing wave pattern, to the left of the NC in Fig. 7 a), or damped out, towards to lower left corner in Fig. 7 b).

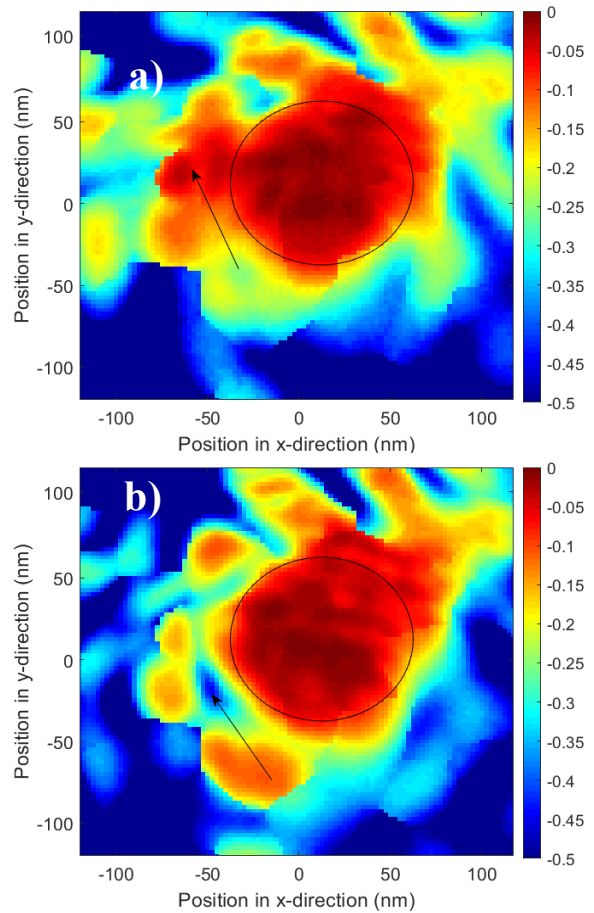

Fig. 7. Zoomed in view of normalized power (dB) at a) the lower and b) higher frequency modes of operation, corresponding to 0-50 ns and 50-100 ns in the simulation. Arrows indicate example standing wave patterns of regions where the oscillation is damped out due to spin wave interference.

In an effort to isolate the dominating reflections in our particular grain tesselation configuration, we identified boundaries or junctions with significantly reduced exchange coupling in close vicinity of the NC area, see Fig. 8. Simulations were done for cases, where the exchange was restored to 100% at the grain boundaries given by the points indicated by letters A-F.

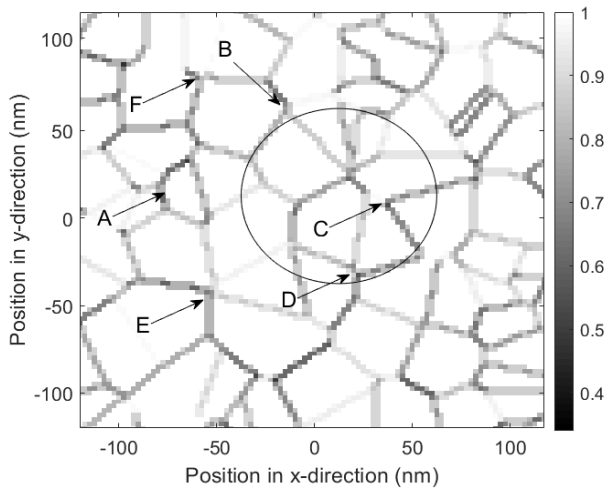

Fig. 8. Zoomed in view of the randomized grain boundary exchange coupling with letters A-F indicating locations of particular interest. The nano-contact is displaced 12.5 nm from the center in the x- and y-directions.

The results is Fig. 9 indicate that the non-linear behavior is governed by the collective effect of spin wave scattering at points that are in within several wavelengths of the NC including points inside the NC or at its perimeter. While progressively restoring the exchange in the grain landscape non-linear effects are still visible, albeit becoming increasingly continuous and shifted towards higher drive currents. The strong reflection at point "E" seems to be mainly responsible for the occurrence of the RTN behavior, since by setting the exchange coupling to 100% at this point the device goes to a continuous non-linear behavior, similar to the nominal case.

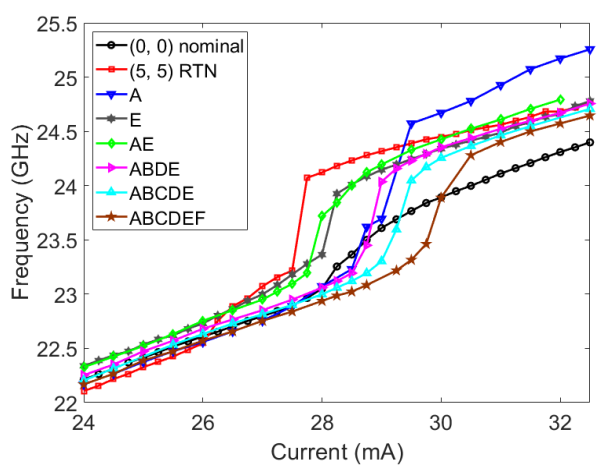

Fig. 9. Non-linear and RTN behavior observed when restoring full exchange (100 %) at grain boundaries in selected locations in the grain landscape.

## IV. CONCLUSIONS

The time domain stability of microwave spin torque oscillators has been investigated by systematic micromagnetic simulations including a randomized grain configuration with reduced exchange coupling at grain boundaries and a stochastic thermal field. Telegraphic mode jumping between frequencies was observed in the time domain with characteristic dwell times in the range of 10-100 ns. The frequency separation between quasi-stable energy states of the oscillator was as large as 1 GHz. The oscillating volume was shown to have a different shape governed by spin wave reflections at grain boundaries with reduced exchange coupling. The resulting non-linear behavior and telegraphic frequency mode jumping of the oscillator were shown to be a collective effect of spin wave scattering at different locations within a few spin wavelengths from the nano-contact.

## V. ACKNOWLEDGMENTS

Partial funding by Swedish Research Council through the project "Fundamental Fluctuations in Spintronics," 2017-04196. We thank J. Åkerman for access to the Ragnarok Cluster at Gothenburg. The computational results presented have in part been achieved using the Vienna Scientific Cluster (VSC) in collaboration with J. Weinbub and L. Filipovic. Access to KTH PDC was granted through Swedish National Infrastructure for Computing.